# A Colour-Encoded Nanometric Ruler for Axial Super-Resolution Microscopies[£,$,¥]

Ilya Olevsko,[a,b] Omer Shavit,[a,b,c] Moshe Feldberg,[b] Yossi Abulafia,[b] Adi Salomon,[a,b,c,✉,*] and Martin Oheim,[c,ID,,]

[a] Department of Chemistry, Bar-Ilan University, Ramat-Gan, 52900 Israel, https://ch.biu.ac.il/;

[b] Institute of Nanotechnology and Advanced Materials (BINA), Bar-Ilan University, Ramat-Gan, 52900 Israel, https://nano.biu.ac.il/;

[c] Université Paris Cité, Saint-Pères Paris Institute for the Neurosciences, CNRS, 45 rue des Saints Pères, F-75006 Paris, France, https://sppin.fr.

[*] co-last authors

[ID] ORCID ID: orcid.org/0000-0001-8139-167X (MO); orcid.org/0000-0002-5643-0478 (AS)         [*]

[✉] corresponding author at

> Prof Adi Salomon
> Department of Chemistry, Bar-Ilan University
> Institute of Nanotechnology and Advanced Materials (BINA)
> Ramat-Gan, 52900 Israel
>
> E-mail: adi.salomon@biu.ac.il
> Lab phone: +972-3-7384236

ABSTRACT. Recent progress has boosted the resolving power of optical microscopies to spatial dimensions well below the diffraction limit. Yet, axial super-resolution and axial single-molecule localisation typically require more complicated implementations than their lateral counterparts. This is, in part, due to the lack of suitable calibration tools that would permit a metrology along the microscope's optical axis. Similar test samples would be needed for quantifying nanometric drift, or axial fluorophore mobility, or again for quantifying the sub-wavelength light confinement in near-field microscopies. In the present work, we propose a simple solution for axial metrology by providing a multi-layered single-excitation, dual-emission test slide, in which axial distance is colour-encoded. Our test slide combines on a standard microscope coverslip substrate two flat, thin, uniform and brightly emitting fluorophore layers, separated by a nanometric transparent spacer layer having a refractive index close to a biological cell. The ensemble is sealed in an index-matched protective polymer. As a proof-of-principle, we estimate the light confinement resulting from evanescent-wave excitation in total internal reflection fluorescence (TIRF) microscopy. Our test sample permits, even for the non-expert user, a facile axial metrology at the sub-100-nm scale, a critical requirement for axial super-resolution, as well as near-surface imaging, spectroscopy and sensing. (201 words).

**HIGHLIGHTS**

- measuring axial fluorophore position is more difficult than lateral localization

- by colour-encoding fluorophore height, we translate a nm-axial metrology challenge into a spectral detection problem

- Electron-beam vapor deposit of $MgF_2$ permits the controlled nano fabrication of thin, transparent spacer layers

- using a dual-color emitting nano-sandwich we estimate the light confinement of the evanescent wave upon total internal reflection

## Graphical Abstract

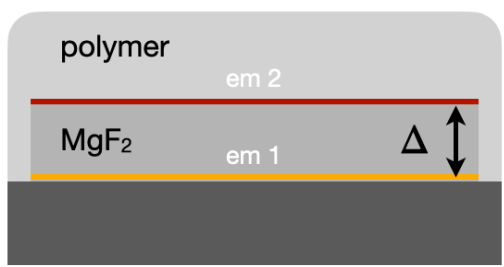
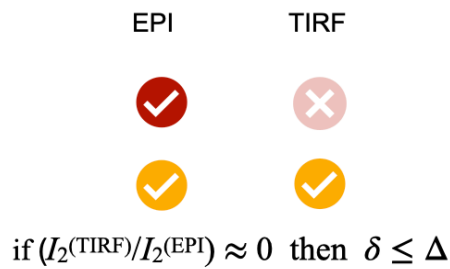

*LEGEND TO GRAPHICAL ABSTRACT*

*Right*, schematic illustration of the nano-fabricated dual-color sandwich, Two thin emitter films (em1, em2) having overlapping excitation, but distinct emission spectra, are separated by a $Mg_2$ spacer of known thickness (here $\Delta$ = 86 nm). This sandwich is covered by MY-133-MC, a non-fluorescent, transparent polymer with water-like optical properties. *Left*, emission matrix of the built sandwich, upon epifluorescence (EPI) and total internal reflection fluorescence (TIRF) excitation, respectively. While the intensities of em1 and em2 are equilibrated upon epifluorescence excitation, the axial optical confinement of the evanescent wave generated upon TIR suppresses emission of the remote fluorophore layer. Depending on the fluorophore height $\Delta$ and the evanescent-wave penetration depth, $\delta$ the TIRF/EPI intensity ratio is a sensitive figure of merit for the light confinement in TIRF.

# 1. INTRODUCTION

## 1.1 Quantifying axial super-resolution and localisation techniques

The lateral resolution of an optical system is often better than its axial one. This is a consequence of the 'missing cone' of spatial frequencies of light along the optical axis[1,2]. Super-resolution microscopies are not an exception to this rule, and despite impressive lateral resolution gains, it has remained challenging to obtain large axial resolution improvements. Often, additional means are required, as illustrated, e.g., by the combination of stimulated emission depletion (STED) for lateral resolution enhancement with surface microscopies like total internal reflection (TIRF)[3,4] or super-critical angle (SAF) fluorescence detection[5] for improved axial sectioning.

Various strategies exist for improving the resolution along the microscopes optical axis, (*i*), increasing the effective NA; (*ii*), multi-view approaches; (*iii*), interferometric techniques; (*iv*) techniques confining the fluorescence excitation or readout volume; (*v*), point-spread function engineering, among others[6]. These approaches share the need to hold the obtained resolution improvement against some external standard. While parallel actin filaments, DNA-origami or sub-diffraction beads provide suitable test samples for lateral metrology, quantifying axial resolution has been more difficult[1,7]. The use of wide-field detection as in the localisation-type microscopies, add additional constraints as the resolution ideally must be quantified across the field-of-view and not only in a single point along the optical axis. What is more, particularly biological super-resolution imaging can be marred by insufficient data quality or artefacts. It is therefore essential to also have biologically relevant control samples to benchmark and optimize the imaging system, labelling and experiment conditions[8]. Curiously, there is a lack of metrological tools for assessing axial resolution[9,10] .

## 1.2 Metrology along the microscope's optical axis

In the present work, we present a resolution target for axial metrology. Our test sample consists of two thin dye layers, axially separated by a transparent spacer layer having a refractive index (RI) close to that of a biological cell. The originality of our sample is that it encodes axial distance by colour[11], thereby converting an axial metrology challenge to a spectral (or multi-channel) imaging and un-mixing problem.

We opted for $MgF_2$ as a transparent spacer material having a suitable refractive index (1.39), which has the advantage of being considerably easier to work with than the earlier used polymer MY-133-MC[12,13] . We identify properties that make a fluorophore pair suitable for dual-emission sandwiches. The chosen fluorophores both excite at 488 nm but they emit distinct deep-orange (peak at 580 nm) and deep-red (720 nm) fluorescence, respectively, allowing for a sensitive detection removed from instrument, oil and glass autofluorescence.

As an example, we here apply such a dual-colour sandwich sample to estimating the light confinement upon evanescent-wave (EW) excitation in an objective-type TIRF geometry[14]. TIRF metrology is not an easy task[10]. Although TIRF intensity measurements from known axial fluorophore distributions allow - in principle - the measurement of the EW-decay, such experiments require either multiple test samples featuring different fluorophore heights[13], **Fig. 1**A (*left*), or moving from one sample region to another, as in the case of the polymer spacer staircase[12], **Fig. 1**A (*right*). In either case, re-focusing or coverslip tilt errors can compromise the accuracy of the measured EW penetration depth.

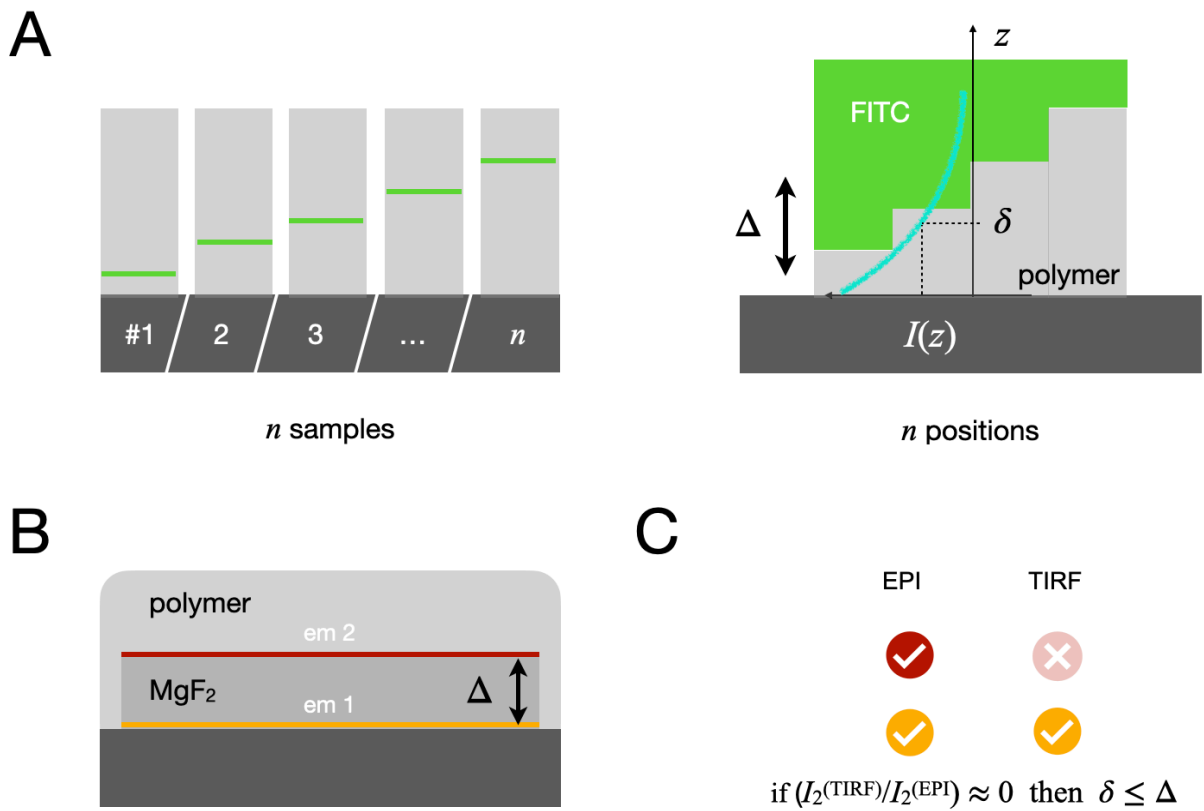

**Figure 1**. *Characterising the evanescent-wave decay by far-field optics.* Different recent test samples for characterising the evanescent-wave (EW) penetration depth in total internal fluorescence (TIRF) microscopy are shown, (A), *left*, series of samples, in which a thin fluorophore layer (*green*) is sandwiched between a transparent, index-matched fluorohpore (*grey*), see Klimovsky *et al.* (2023). (*right*), similar strategy but using a polymer 'staircase' obtained by successive dip-coating and topped with a dilute fluorescein (FITC) solution Niederauer *et al.* (2018). $\Delta$ and $\delta$ identify, respectively, the fluorophore height and EW decay length. Turquoise intensity decay, $I(z)$, schematizes EW decay. *Dark grey* rectangle is borosilicate glass substrate. (B), scheme of the here proposed new calibration sample for axial distances, featuring two emitter layers (em1, em2), separated by a nm-thick spacer layer made from $MgF_2$ and capped with a thick polymer layer for longer shelve-life. em1 and em2 are both excited by 488-nm light but emit at different colours, *deep red* and *orange*, respectively. (C), fluorescence matrix upon 488-nm epifluorescence (EPI) and TIRF excitation, respectively. For the test slide shown in (B), the absence of red indicates an evanescent-wave penetration depth $\delta$ smaller than the fluorophore height, $\Delta$.

Our color-multiplexed sandwich samples allow for a 'one-shot' measurement. In the example presented here, the two emitter layers are separated by 87 ± 4 nm, making it a convenient axial ruler on the length-scale probed by total internal reflection fluorescence (TIRF) or supercritical angle fluorescence (SAF) microscopies, **Fig. 1***B*. Upon epifluorescence excitation (EPI) both dyes fluoresce, but the emission of bottom layer dominates when switching to TIRF. Our test sample thus permits the facile assessment of the excitation confinement using dual-band- or spectral detection, **Fig. 1***C*.

Our paper is organised as follows: section 2 assembles the relevant experimental procedures, before, in section 3.1, we present a method for the controlled deposit of transparent, ultra-thin, homogenous, and flat films of $MgF_2$ having a RI close to that of a biological cell. We rationalize in section 3.2 the choice of the fluorophore pair, TPPS J-aggregate and Rubpy. Section 3.3 describes the assembly of the nano-layered sandwiches that we subsequently image on an objective-type TIRF microscope to probe the axial sectioning at various beam angles, using a red/orange emission ratio (section 3.4). We conclude with a discussion and perspectives in section 4.

## 2. MATERIALS AND METHODS

Additional experimental procedures are found in the Supporting Information Online.

### 2.1 Substrates

We used optical-grade borosilicate coverglasses (#1, 25-mm diameter, Menzel-Gläser, Braunschweig, Germany) as a substrate. Coverslips were thoroughly cleaned through a multi-step process in ethanol, distilled deionised water (DDW), and a commercial alkaline cleaning concentrate (Hellmanex, Hellma Advanced Optical Components, Jena, Germany, see Fig. S1 and SI Materials). Cleaned substrates were either used immediately, or else stored in a clean and sterile Falcon tube.

### 2.2 $MgF_2$ spacer layers

Transparent spacer layers were deposited by electron beam physical vapor deposition (e-beam PVD) on a custom electron beam evaporation system (BesTec, Berlin, Germany, for BINA Nanocenter) with an acceleration of 9.5kV, under vacuum (~$10^{-5}$ mbar) and with stage rotation at 5 rpm and no stage heating. We monitored the growth of the layer in real time by a film-thickness measurement (FTM) device. For the deposition of spacer layers on glass substrates we used 1-6 mA current, providing a growth rate of 1-3 Å/s. When depositing the spacer layer onto a previously deposited emitter layer we employed more gentle conditions: the e-beam was focused on the target with the shutter closed (which prevents the $MgF_2$ vapor to hit the substrate) during at least 40 s to clean the $MgF_2$ target surface from any contaminations. Only then, the shutter was opened with a work current of 0.5-1 mA to provide a slow growth rate (~ 0.1 Å/s) of the first 10-20 nm. Thereafter, the current and rate were raised to normal values. We alternatively explored the deposit of $MgF_2$ layers by ion beam sputtering (IBS) using an Intlvac Nanoquest I (Georgetown, Canada) at a vacuum base pressure of ~$10^{-6}$ mbar, a flow of 10 atm $cm^3$/min (sccm) and 110 mA ion beam current at 1,2 kV voltage (see SI Materials).

**2.3 Solutions**

TPPS J-aggregate (4,4′,4″,4‴-Porphine-5,10,15,20-tetrayl tetrakis benzenesulfonic acid) was purchased from Sigma-Aldrich (#88074). A 0.57-mM stock solution was prepared by dissolving 11.6 mg TPPS powder in 20 ml $HNO_3$ aqueous solution with pH<1. Tris (bipyridine) ruthenium (II) chloride ($[Ru(bpy)_3]^{2+}$ $2Cl^-$) powder was purchased from Sigma Aldrich (#544981). 150 mg of Rubpy powder was dissolved in 2 ml DIW, and aqueous dilutions were prepared from the 100-mM stock solution. Poly-diallyl-dimethylammonium-chloride (PDDA) stock solution (35% wt/wt in water, 100 kDa MW) was purchased from Sigma Aldrich #522376) and a 1:10 dilution (DIW) used as a primer layer onto which we deposited the fluorophore layers.

**2.4 Thin-film deposition**

Thin molecular films were produced using a modified layer-by-layer (LBL) technique that required only a few tens of µl of dye solution and allowed us to deposit the film on a specific area of the sample with minimal spill elsewhere. This was achieved by liquid deposit, rather than dipping the sample into the solution. This procedure protected the $MgF_2$ spacer layer and multiple fluorophores could be deposited using the same approach, with minimal modifications, and only on one side of the substrate. Briefly, the substrate was glued onto a standard microscope slide with stripes of Capton® tape (polyimide film). Then, the surface preparation treatments are performed, (*i*), Corona surface activation (CT, also referred to as plasma activation, on a iCorona TF-415 (Vetaphone, Kolding, Denmark), 40-45 s at 1kW power; (*ii*) a PDDA primer layer was then deposited onto the surface to better attach the negatively charged dye molecules. We note that optimal thin film-deposition conditions had to be developed for every dye. The different solutions were drop-cast onto the samples surface using a μl Gilson pipette. The volumes applied covered an area of ≈1-cm diameter. Excess solution was blown away with a $N_2$ gas steam that also served to dry the surface[15].

*TPPS films* - on a BK-7 substrate, 20 μl of PDDA solution were applied and blown away after 30 s. Then, 20 μl TPPS solution were drop-cast, again removed after 30 s with an $N_2$ flow. For the deposition onto $MgF_2$ 20 μl of dye solution was cast without surface treatment, and subsequently blown away.

*Rubpy films* - following CT, 10 μl dye solution were drop cast on the surface and removed after 20 s with $N_2$ flow. Following dye deposition, the sample was cleaned to remove any traces of dye on the far side of the substrate. Samples were stored in desiccator with dehydrate silica beads, under low pressure (0.013 atm) at room temperature (RT, 22-23°C) during at least 48 h.

*Capping layer deposition* - MY-133-MC polymer (MY Polymers LTD, Ness-Ziona, Israel) was applied by drop casting the undiluted stock solution on top of the dry sample surface. For curing the samples were stored under a wet beaker overnight[16]. The cured polymer formed a thick (>2 μm) transparent layer that provides a water-like optical environment (RI = 1.33-1.34, [13]) and protects the dye and spacer layers, which notably allows for multiple uses of the same calibration sample.

**2.5 Thin-film characterisation**

Spacer layers were characterised in terms of their thickness, roughness and homogeneity using a combination of Stylus profilometry, ellipsometry, atomic-force microscopy (AFM). The AFM images were used to measure peak heights and to calculate the average roughness ($R_a$) of the deposited layers.

*Dark-field microscopy* - Sample homogeneity and fluorophore crystals/aggregates were monitored on an upright reflection dark-field microscope (BX51, Olympus, Hamburg, Germany) using a ×20/ NA 0.45 objective, and Olympus camera and program.

*Transmittance and fluorospectrometry* - we used an inverted microscope (IX83 Olympus) equipped with a 40×/0.6NA objective, linked to an imaging spectrometer (IsoPlane SCT320, Teledyne Princeton Instruments,Trenton, NJ), equipped with a 600-nm blaze/50 grooves/nm grating and an electron-charge coupled device detector (EMCCD, PIXIS 1024 eXcelon, TeleDyn PI). The setup was routinely calibrated using Princeton Instruments IntelliCal® calibration device. Transmittance spectra were obtained upon broadband halogen illumination (100 W) using 10-ms exposure including an IR filter (750LP). Transmitted light was collected at 150-µm slit width, and the absorbance A = $\log(1/T)$ = $\log(P_0/P)$ calculated from the measured transmission spectrum ($T(\lambda)$). For fluorescence spectroscopy, the excitation light provided by a Xe-arc lamp (85% power) was filtered with a narrow 488-nm band-pass filter (ZET488/10x BP, AHF Analysentechnik, Tübingen, Germany) and a matched dichroic mirror (T495LPXR). Emission spectra were collected at 250-nm slit width through a 500LP filter for spectral acquisitions. Specific BP filters were used where indicated. The shown spectra are the average over 1024 pixel rows, see Fig. S2.

*Back-focal plane (BFP) imaging* for the detection of supercritical (SAF) and under critical (UAF) angle fluorescence used a home-built inverted micro-scope assembled from optical bench components, see Klimovsky et al. (2023)[13] and Fig. S3 in the Supporting Information Online. Briefly, the beam of a 488-nm laser (Coherent Sapphire SF 488-50) was cleaned up with a 488-nm notch filter, the polarization turned to perpendicular, the beam attenuated with neutral density filters, spatially filtered and expanded to 2" beam diameter. It was then scanned by a tip-tilt mirror and tightly focused in the BFP of a ×100/1.46NA TIRF objective (αPlan-Apochromat, oil DIC M27, ZEISS, Oberkochen, Germany). Images were acquired upon EW excitation (with the spot positioned in the extreme periphery of the objective's BFP) or EPI excitation (spot on-axis, at 0, 0). We measured the maximal effective power at the sample plane as 0.6 mW in EPI using a PM100D powermeter with S130C (400-1100 nm) photosensor (Thorlabs). Fluorescence was collected through the same objective, filtered by a ultra-flat (2-mm) long-pass 488LPXR dichroic mirror (AHF), RET493LP (Semrock, IDEX health science, Oak Habor, WA) and ET520LP (Chroma, Bellows Falls, VT) long-pass filters (see again Fig. S2). BP filters were housed in two stacked independent filter wheels as stated in the figure legends, see Table S1 in the Supporting Information Online for a complete listing). The filtered fluorescence was imaged on a back-illuminated sCMOS camera (edge4.2bi, PCO, Kelheim, Germany). The effective pixel size was 30 ± 1 nm/pixel for sample-plane (SP) images, allowing for a 3- to 4-fold binning without resolution loss. A Bertrand lens mounted on a motorized flipper (Thorlabs) allowed us shifting the focus from the SP to the objective's back-focal plane (BFP). The pixel size was 3.36 µm/px for BFP images.

*Dual-band emission ratiometry.* Sample-plane images were subtracted with a dark image and the intensity of a region of interest on the resulting image averaged. Intensities measured upon 488-nm EPI or TIRF exciation through a BP600/25 emission band-pass (for Rubpy) or BP720/60 (for TPPS) were used to calculate the emission ratios, $I_{TPPS}^{(j)}/I_{Rubpy}^{(j)}$ or $I_{Rubpy}^{(j)}/I_{TPPS}^{(j)}$, where the index j indicates the illumination mode, EPI or TIRF, respectively.

*BFP-image analysis.* The ratio of the integrated supercritical (SAF) vs. undercritical angle fluorescence (UAF) emission is a direct, illumination- and concentration-independent measure of fluorophore height [10,13,17]. BFP images were analyzed using MATLAB script (The MathWorks, Natick, MA). First, the background was subtracted by using the corresponding dark image to remove intensity offset and light contaminations. Then the BFP was segmented to three zones in a multi-step process: (*i*), the centre (optical axis) was found on the BFP image by binarising the image and fitting a circle with the circular radiation pattern from a pyranine (text marker highlighter pen) sample. (*ii*), then, the outer radius, corresponding to the limiting NA ($r_{NA}$) was found. This measurement gives the effective NA ($NA_{eff}$) of the very objective used ($NA_{eff}$ = 1.465 in our case, [13]), which is generally slightly different from the manufacturer-specified NA (see also [18–20]). Finally, (*iii*), the emission critical angle of the radiation pattern (related to $r_c$ via $r_c = f_{obj} n_1 = r_{NA} n_1 / NA_{eff}$)) was calculated based on a the independently measured sample RI ($n_1$), see above. Error bars were generated according to $r_c^{(\pm)} = r_{NA} \times (RI \pm dRI)/NA$. Based on the obtained parameters (centroid, $r_{NA}$, $r_c$) the BFP image was segmented to three areas: background, SAF and UAF. The intensity was integrated over the SAF and UAF regions, and the SAF/UAF intensity ratio calculated, $R = I_{SAF}/I_{UAF}$, where I are the respective integrals). Repeating this integration for each dRI gave us an estimate of the dependence of SAF/UAF ratio on the accuracy of the RI estimate. For this work, analysis was performed with RI=1.394 with dRI = ±0.02 for the error bars.

## 3. RESULTS

### 3.1 Fabrication of transparent nm spacer layers by MgF$_2$ electron-beam deposition

A key requirement for a TIRF-intensity calibration sample is a precisely controlled axial fluorophore distribution above the reflecting interface[10]. We aimed at a thin, flat and homogeneous emitter layer that covers a large surface area, at least the field of view of a ×60 microscope objective (~0.5 mm, the smallest magnification available for TIRF-capable objectives). These features will considerably facilitate the use of the calibration samples for routine metrology and instrument troubleshooting. Also, both the spacer and capping layers (see Fig.1) should have a RI close to that of a biological cell (1.360-1.391 see, e.g. Ref 12) to provide measurements relevant for biological TIRF imaging. In our earlier work, we used a transparent non-fluorescent polymer spacer (MY-133-MC)[13]. Systematically varying the polymer concentration and the rotational frequency of the spin coater, we could consistently control the thickness of these spacer layer, but we found MY-133-MC somewhat awkward to work with. We therefore explored alternative strategies for fabricating transparent, low-index spacer layers. Electron-beam (eBeam) evaporation of magnesium fluoride ($MgF_2$) is known for antireflective coatings[21]. We here employed a similar strategy for producing transparent layers of controlled nm thickness.

$MgF_2$ is relatively inexpensive, chemically stable, transparent in the visible, and it has a suitable RI for our application, ~1.39 at 490 nm[22]. We fabricated smooth and homogenous $MgF_2$ films on cleaned BK-7 glass substrates, **Fig. 2***A* by Ebeam vapor deposition, **Fig. 2***B*. The resulting layers displayed a remarkable uniformity and low roughness. The measured average roughness ($R_a$) was typically <1.5 nm **Fig. 2***C* and Fig. S4. Ellipsometry revealed a RI of 1.39(4) for Ebeam-deposited $MgF_2$ layers at 490 nm, Fig. 2D (*green*), slightly higher than the value reported for crystalline $MgF_2$ (RI =1.38). This discrepancy could result from some residual poly-crystallinity in our layers [22]. On the other hand, $MgF_2$ layers that were deposited by ion beam sputtering (IBS) showed RI values of ~1.43 at 490 nm **Fig. 2***D* (*red*), too high for our application. Of note, the fit of the crystalline $MgF_2$ model with the ellipsometry data was quite poor (MSE > 80 vs. 30 for the Ebeam data) indicating that IBS-produced $MgF_2$ layers were not ad-equately described as a $MgF_2$ crystal. A better fit could be achieved with the B-Spline model (*data not shown*). This might indicate a stoichiometric defect [23] resulting from the high energy involved in IBS deposition.

Together these experiments indicate that Ebeam-deposited $MgF_2$ layers, with their RI value close to that of a biological cell and their mono-crystallinity, are well-suited for RI-matched spacer layer fabrication.

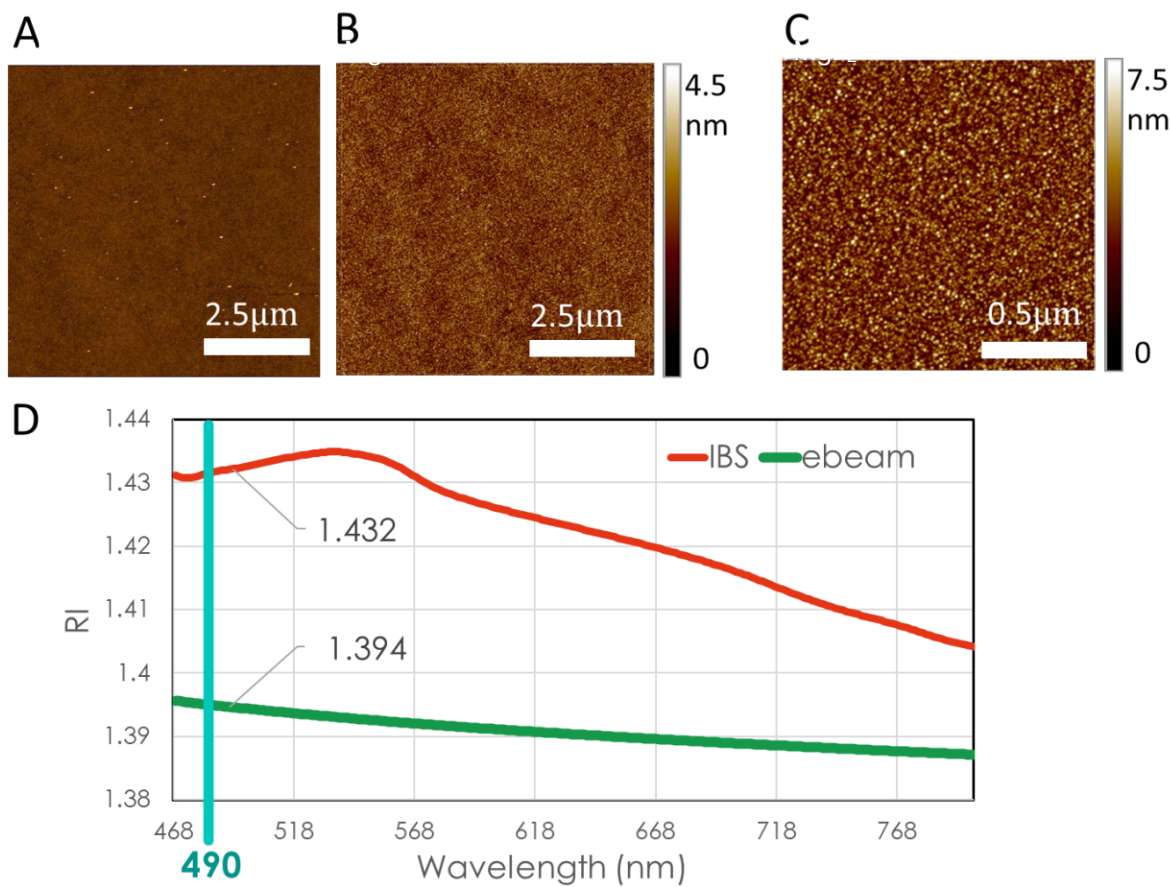

**Fig. 2.** *Physical characterization of the $MgF_2$ spacer layers.* (A), AFM example image of a nude borosilicate (BK7) coverslip, following surface cleaning (see Methods and Supporting information). (B) AFM example image of a 100-nm thick $MgF_2$ layer deposited on a cleaned BK-7 substrate. Height scale is the same for (A) and (B). (C), zoomed view of (B). (C), ellipsometric measurements of refractive index (RI), as function of wavelength, for two 100-nm $MgF_2$ layers deposited either by E-Beam (*green*) or IBS physical vapor deposition (*red*), respectively.

### 3.2 Brightness- and spectrally-matched dyes for dual-colour sandwichs

Disposing of transparent spacer layers of low roughness and high flatness with a RI close to that of a biological cell, we next searched for a fluorophore pair that offers overlapping excitation but well-distinct fluorescence emission spectra, permitting the colour discrimination of the layers, either in a dual-band emission detection scheme, or via spectral imaging. As an excitation wavelength, we opted for the popular 488-nm line of an $Ar^+$ or diode laser that is used in many TIRF experiments.

We used TPPS J-aggregates that form thin films of sufficient brightness and emit in the far-red/NIR, We first investigated the physical properties of the previously used TPPS J-aggregates when deposited on the Ebeam-deposited MgF$_2$ surface. TPPS J-aggregates are deep-red emitting chromophores (720 nm peak upon 490-nm excitation[24,25] allowing for an easy discrimination against immersion oil and instrument autofluorescence (AF). The aggregate height is about 5 nm[24], small enough for thin-layer formation, but they tend to form larger bundles under acidic conditions[15], **Fig. 3***A*. Due to the requirement for thin and smooth dye layers for our nanoruler fabrication, we aimed to establish conditions for dense J-aggregate surface-coverage without bundle formation. For MgF$_2$ spacers, we directly drop-cast the dye solution was on the CT-cleaned surface, **Fig. 3***B*, while for the fabrication on TPPS films with similar aggregate density on BK-7 substrates ("zero"-samples), a pre-coating with a thin PDDA polymer layer was required (CT followed by drop-cast of 3.5% solution of low molecular-weight PDDA chains). We explain the higher affinity of TPPS J-aggregates to the MgF$_2$

surface due to the presence of negative charges (fluoride ions), and to the higher roughness of the MgF$_2$

layers compared to BK-7 (see above).

We formed uniform TPPS layers on either substrate, consisting of well-separated, small aggregates. Dark-field and fluorescence images revealed a dense and uniform dye coverage of cm$^2$ surface areas. The observed negligible emission at 670-nm and low 645-nm absorbance peaks (associated with the TPPS monomer[25]), confirm the predominance of J-aggregates. This is plausible, because the negatively charged monomers have a lower affinity to the substrate compared to the neutrally-charged aggregates, which are zwitterions. AFM measurements, **Fig. 3***D*, showed an average $R_a$ < 2 nm for layers of 5-nm thickness, with no presence of bundles. Higher TPPS concentrations (>1mM) or stronger acidity (pH<1) produced layers that were a mixture of both single aggregates and bundles. Under these conditions, the bundle height exceeded 20 nm (panel A, *left*). Although the various shapes of those rods can be of use to find the focal plane, these samples were less suited than the thin J-aggregate layers for our application because of local intensity variations and a less defined axial fluorophore position.

As the second fluorophore, we opted for a Tris (bipyridine) ruthenium (II) (Rubpy) complex [Ru(bpy)$_3$]$^2$ Cl$^-$ (Rubpy). Our rationale was to find a dye having a shorter emission than TPPS, but still in the red, removed from glass [26,27] and immersion-oil autofluorescence [28]. Rubpy emits around 600 nm and it can be exited in the 430-490 nm range, [29], compatible with the used Ar$^+$-laser line. In addition, Rubpy is photostable and it can form thin fluorescent films [30].

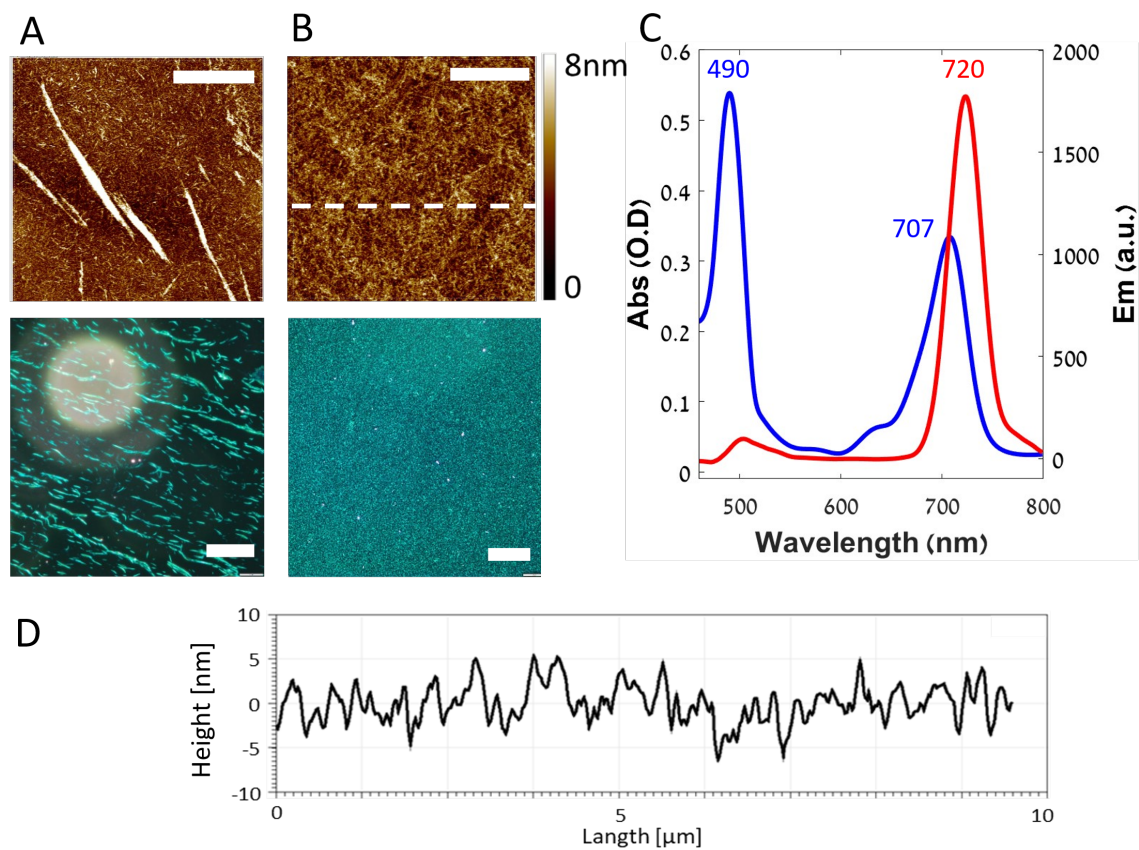

**Fig. 3.** *Characterization of TPPS J-aggregate dye layers deposited on top of a 192-nm thick MgF$_2$ spacer layer.* (A), Images of a TPPS dye surface prepared under very acidic conditions (1 mM, pH<1). Note the bundles seen on the AFM image (*top*) and the corresponding dark-field (DF) image (*down*) (B), same, but for 0.5 mM, pH=1.2. Note the absence of fluorophore bundles and the more uniform distribution (C), absorbance (*blue*) and fluorescence emission (*red*) spectra, respectively, of the TPPS J-aggregate layer presented in (B). (D) AFM line profile measured along the 10 μm dashed line on the thin TPPS film presented in (B). Scale bars are 2.5 μm for AFM scans and 40μm for DF images, resepectively.

Deposition of Rubpy results as a thin layer of Rubpy ~ 30 nm crystals, **Fig. 4***A*. The layer formation is guided by electrostatic interaction between the negatively charged CT-treated surface and the positively charged dye complex. Shorter exposures of the surface to the solution (20 s) or lower dye concentrations (e.g., 10 mM, **Fig. 4***B*) prevented the formation of big crystals. Instead, the nucleation was mainly on the surface where thin flakes formed. Films obtained from a 10 mM solution showed a dense and homogenous coverage with tiny crystals and with absorbance and emission spectra similar to those in solution, **Fig. 4***C*. AFM confirmed maximal crystal heights of ≈ 30 nm, and a R$_a$ roughness of ≈8 nm. Fluorescence intensities varied <10% over a 4-mm$^2$ area, **Fig. 4***D*.

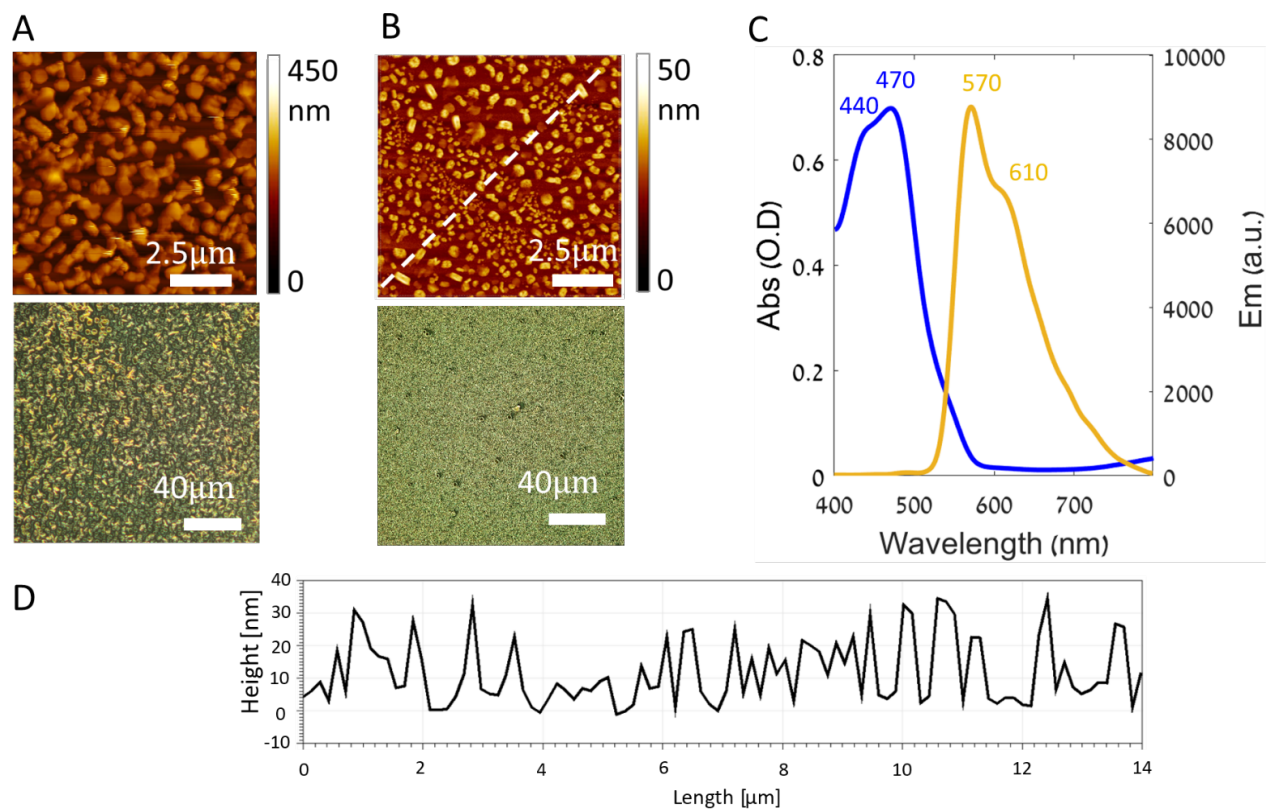

**Fig. 4.** *Characterization of a Rubpy layer deposited onto a 192-nm spacer layer of MgF₂ (A), AFM image (top) of Rubpy layers deposited from a 100-mM dye solution and the corresponded dark-field image (bottom (B), same, for 10-mM dye solution. Note the smaller crystals and reduced height of the fluorophore layer (C), absorbance (blue) and fluorescence (yellow) spectra of the Rubpy films presented in (B). (D), AFM line profile for the Rubpy film shown in (B), measured along the dashed line.*

Lower Rubpy concentrations reduced the density of the coverage, higher ones led to significantly thicker layers and higher $R_a$. For example, the crystals obtained from a 100-mM solution were about 300 nm high, with an $R_a$ of 48 nm and thus unsuitable for our purposes (see **Fig. 4**A (panel 1) for the AFM image). Interestingly, larger crystals displayed a slightly red-shifted emission with a peak at around 610 nm (*not shown*), possibly associated with a higher water content, [31] (in fact, 620 nm is the peak value observed in an aqueous solution). All taken into account, we opted for the smaller and brighter Rubpy crystals prepared from a 10-mM solution for our axial-ruler sandwiches.

**3.3 Axial distance coding by colour multiplexing: dual-colour sandwiches**

In addition to spectral considerations, other parameters govern the usefulness and performance of our dual-colour sandwiches, e.g., the precise fluorophore height and the relative emission $F_{\mathrm{TPPS}}/F_{\mathrm{Rubpy}}$ from the two dye layers. Here, $F_i^{(j)} = I_0^{(j)} c_i B_i Q_i$ and $c_i$ and $Q_i$ are, respectively, the concen-tration of dye i and the quantum yield of the detector in the emission band of this fluorophore. $B_i$ is the earlier defined molecular brightness of fluorophore i. Index j denotes the illumination mode, TIRF or EPI. A compromise must be found to equilibrate these parameters, because the dual-colour sample is going to be illuminated upon EPI as well as in TIRF excitation, and at various beam angles, $\vartheta$, resulting in various penetration depths, $\delta = (\lambda/4\pi) \cdot [(n_2 \sin\vartheta)^2 - n_1^2]^{-1/2}$, and hence various intensities that will be emitted from the two layers. As expected[13], more distant layers were less excited upon EW excitation (see Fig. S5). From these preliminary experiments we decided to equilibrate the signals emitted from either dye layer upon EPI illumination, i.e., $I_{TPPS}^{(EPI)} \approx I_{Rubpy}^{(EPI)}$, but other choices are possible. Dye concentrations and resulting brightnesses could be controlled by the way the layers were deposited (see Supporting Experimental Procedures in the Supplementary Information Online).

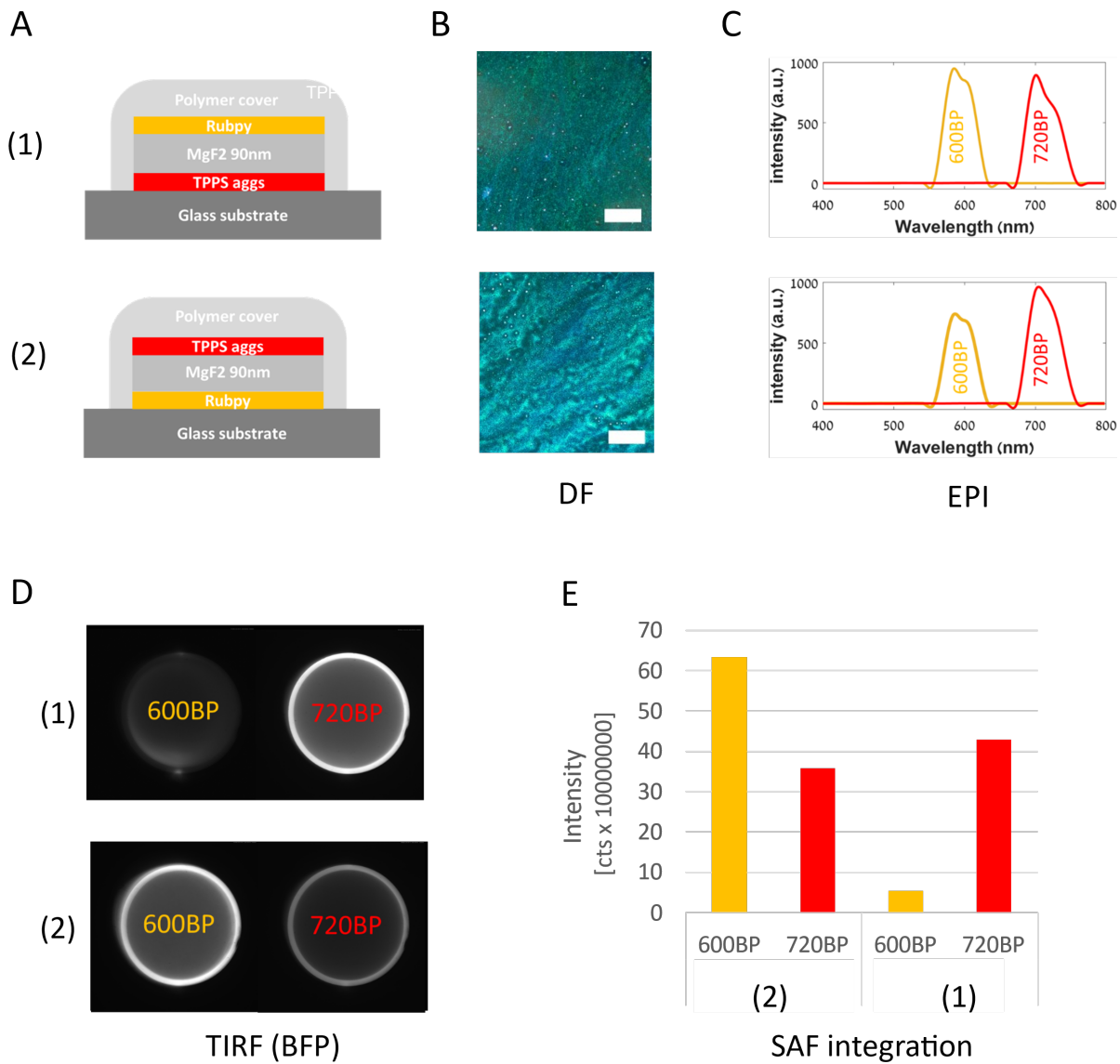

**Fig. 5** *Single-wavelength excitation, dual-color emission sandwich samples.* (A), illustration of the two prepared dual-color sandwiches. (B) Dark-field (DF) images of the TPPS J-aggregates layers: the sample topologies were similar, independent on whether the dye was deposited on glass or on the spacer (scale bar: 40 μm). (C), Effective emission spectra recorded from the samples shown in (A) through band-pass, filters as indicated, upon epifluorescence (EPI) excitation. Color code as in (A). Note the similar brightness for both dye layers. (D) TIRF excitation revealed marked emissionintensity differences, with the respectively more remote dye layer much darker. Images show BFP (Fourier-plane) images that reveal the corresponding radiation patterns and allow for the segmentation of the emitted signal into under-critical and super-critical emission components, corresponding to fluorophores far and very close to the reflecting interface, respectively. Excitation at 488 nm; exposure time 1s. (E) integration of intensities emitted into the supercritical angles (SAF area) of the BFP images presented in D show the expected intensity dependence upon fluorophore height. See main text for details.

### 3.4 Intensity ratiometry as a figure of merit for EW confinement

Assembling the previously characterised building blocks (TPPS and Rubpy layers separated by a MgF$_2$ spacer, respectively) we next fabricated two dual-colour sandwich samples, the sequence of the fluorophore layers being their only difference, **Fig. 5**A. In either case, a 87 ± 4 nm thin MgF$_2$ spacer separated the two emitter layers (height confirmed by profilometry, *not shown*). Like the earlier single-dye layer samples (fig.3 and fig.4, respectively), our nanofabricated dual-colour sandwiches displayed a uniform dye coverage over large surface areas, as seen from the corresponding in-focus dark-field images, **Fig. 5**B. EPI excitation at 488 nm revealed similar emissions $I_{TPPS}^{(EPI)} \approx I_{Rubpy}^{(EPI)}$ irrespective of the precise dye sequence, **Fig. 5**C.

Switching to EW excitation dramatically changed the observed emission intensities, **Fig. 5**D. The observed intensity changes were not the result of selective surface quenching, bleaching or other artefacts, because the corresponding back-focal plane (BFP) images revealed the expected opposite evolution of the fluorophore radiation patterns[13,17], depending on which fluorophore was close or distant to the reflecting interface. In addition to intensity information, BFP images contain a *k*-space representation or angular emission information (radiation pattern) of the collected fluorescence, averaged over all emitters in the field-of-view. Radii in the BFP image can be thought of as equivalents to numerical apertures (NAs) with the outer limit representing the effective NA of the objective lens used, and camera dark noise and background outside. Within the objective pupil, two zones can be distinguished: a thin outer ring (between NA$_{eff}$ and the RI of the imaged sample[17, 32, 37]) that corresponds to supercritical angle fluorescence (SAF) emitted into 'forbidden angles' for light collected from fluorophore located far from the substrate. That SAF is observed at all, is a consequence of the proximity of the emitter to the substrate and the conversion of near-field emission of the dipole to propagating light, detectable in the far-field under extremely high angles otherwise forbidden by Snell's law. The inner zone of the radiation pattern corresponds to the usual ('under critical angle') fluorescence (UAF) emission.

With that in mind, we note that on **Fig. 5***D* both orange (600±25 nm) and deep-red (720±30 nm) colour channels follow the expected radiation patterns. Not only the surface-proximal fluorophores show a brighter emission, but they also display a bright SAF emission component, while the surface-distant fluorophore layer emitted a lower fluorescence and displayed a virtually absent SAF emission component, **Fig. 5***E*. The fact that the results are not completely symmetric upon inverting the two emitter layers is an effect of spectral cross-talk: with TPPS at the bottom (and Rubpy above) a higher contrast between the two dye layers was observed upon EW excitation, while the reverse sequence (Rubpy on the surface, TPPS above) presented a smaller intensity difference and still detectable SAF in the far-red channel. This is plausible, as Rubpy - the shorter-wavelength emitting dye - when at the bottom, is stronger excited as it is closer the interface, and its emission tail leaks into TPPS channel where it contaminates the anyway fainter intensity emitted from the TPPS layer. An additional, but negligible, effect comes form the wavelength-dependence of the dipole's emission near-field, which allows the 720-nm emitters to couple more efficiently to the surface than the 600-nm emitters, and hence also (slightly) increases the SAF emission component of the longer-wavelength dye.

We conclude that equilibration of the EPI intensities is a good starting point for multi-layered sandwiches, but the dye sequence must also take into account their emission spectra and the available filters used. A long-to-shorter wavelength-emitting sequence from bottom to top is generally advisable for optimal performance.

**4. Discussion and perspective**

Calibrating EW penetration depths for biological TIRF microscopy ideally involves test samples that - as closely as possible - mimic the biological sample under study. Also, these samples should feature a known, fluorophore distribution so as to relate the measured intensities unambiguously to axial fluorophore distances [10,13]. Spatial uniformity, stability over time and minimal interference from autofluorescence are additional critera. With the EW decay being the only unknown, the effective EW penetration depth can be recovered from samples with different, known fluorophore heights. Many different techniques have been proposed (see ref. 10 for review) but - at present - there is no perfect solution, and no single technique has been adopted by the community. This is, in part, due to the need for fabricating test samples, their short shelve-live and other limitations. In this context, we aimed at providing a solution for better reproducibility and quality control for TIRF and other axial super-resolution microscopies that is as simple and fool-proof as the green fluorescent plastic "Chroma slides" are for epifluorescence. As a first step in this direction, we recently added to the panoply of axial rulers one[11] which stands out by providing a simple nano-layered architecture on a large spatial scale[13]. Reminiscent of thin films, optical cavities or etalons, our sandwich samples feature multiple layers of tightly controlled nanometric thickness and flatness, separated by a transparent, index-matched spacer and sealed with a capping polymer, both having a RI close to that of a biological cell and thus creating a powerful test sample for TIRF microscopies and related techniques like supercritical angle fluorescence[32–37] (see ref. 17 for a recent review), surface-plasmon resonance, non-radiative excitation fluorescence microscopy with Qdot emitter layers [38] or metal-induced energy transfer (MIET)[39].

Ideally, such axial calibration samples would include fluorophore monolayers that with their molecular thickness offer the most precise fluorophore localisation, but such thin layers are too dim and bleach rapidly. We therefore sought a compromise between detectability and robustness on the one hand, and axial localisation precision on the other.

Furthermore, inn the current work we built on our earlier approach of nanoscopic sandwiches[13] and we extended the idea to multiple spectrally different fluorophore layers that allow for an colour-encoding of axial fluorophore distance (fig.1). We also improved the spacer material compared by the use of e-beam $MgF_2$ vapor deposition (fig.2). We characterised the physico-chemical properties of two red-emitting fluorophores when deposited as thin films, TPPS (fig.3) and Rubpy (fig.4). Assembling dual-colour nano-sandwiches from these components we show their use in a single-wavelength excitation, dual-colour detection scheme (fig.5).

### 4.1 TPPS J-aggregate films: homogenous and thin, but dim

We produced and imaged 4- to 5-nm thin, smooth ($Ra \approx 2$ nm) TPPS J-aggregate films. Similar to our earlier work[13] thin layers of TPPS aggregates were deposited at well-defined axial distances from the glass substrate, using a transparent spacer layer (see below). The main disadvantage of TPPS J-aggregate layers is their still relatively low brightness. Despite a high extinction coefficient ($\varepsilon = 350{,}000$ $M^{-1}cm^{-1}$) the low quantum yield ($\phi_F = 0.03$) [25] results in an overall dim fluorescence, which is also true for thin films. Therefore, with the here used several 100 μW excitation powers, long exposure times of the order of 1s were required, even with high-NA detection and highly-sensitive detectors.

### 4.2 Rubpy: crystalline, spectrally well-separated but difficult large-scale homogeneity

Modified drop-LBL produced Rubpy films showed a peak absorption at 470 nm with a shoulder at 440 nm, and a fluorescence emission maximum at 570 nm with a shoulder at 610 nm. Rubpy crystals had higher emission compared to solid films. However, the resulting films where thicker compared to TPPS aggregate layers, of the order of tens of nm compared to <5 nm for TPPS.

Likewise, the surface roughness was about an order of magnitude higher (approximately 10 nm for Rubpy vs. 1 nm for TPPS), probably due to the crystalline nature of surface-deposited Rubpy. While this crystalline morphology facilitates focussing compared to the uniform TPPS films, but - together with the local concentration, and hence brightness, variations - the resulting Rubpy layers are not ideally suited producing thin and uniform films over a large scale.

**4.3 MgF$_2$ as a spacer material.**

Magnesium Fluoride (MgF$_2$) produces a highly pure, dense material form that is particularly well suited for optical coating. As a low index material, MgF$_2$ has been used for many years in anti-reflection and multilayer coatings. It is insoluble and hard when deposited on hot substrates. Its optical, chemical, and mechanical properties make MgF$_2$ suitable for our application. Vacuum as well as nonvacuum techniques are suitable for its production, including chemical aerosol decomposition of magnesium-fluoro compounds, as well as spinning or dipping into solutions of the magnesium-fluoro compounds[40,41]. Reported uses of MgF$_2$ as a spacer material on the other hand are rare. One recent study optimised Surface Plasmon Resonance (SPR) sensors by using MgF$_2$ layers. In the Kretschmann configuration, the prism surface was first coated with a smooth 50-nm gold film, followed by a MgF$_2$ layer as a transparent spacer. To obtain the maximum reflection dips in SPR, the authors varied the thickness of MgF$_2$ layer[42].

In the present study, we made a similar use of ultra-thin MgF$_2$ layers as a transparent spacer material. We used e-beam PVD to produce thin MgF$_2$ films of precisely controlled thickness. Compared to the earlier used MY-133-MC layers, we found MgF$_2$ deposits easier to control, free from impurities and air bubbles and having a longer shelve life while still providing a transparent spacer of a RI close to that of a biological cell.

**4.4 What's next?**

Future work must address several shortcomings of the current procedure and samples:

(*i*), the low brightness of the used TPPS layers and also the large relative difference in brightness with TPPS are a problem, which result in long exposure times, but also spectral cross-talk. Clearly, different fluorophores having better photophysical properties are desirable.

(*ii*), the spatial inhomogeneity of the dye concentration across the films because of crystal formation and uneven surface coverage: although <10% in the worst case, this leads to a space-dependent intensity imbalance between the emitters and introduces error and uncertainty;

(*iii*), the large surface roughness $R_a$ and thickness of Rubpy layers ($\approx$30 nm) is a disadvantage for their application as an precise nano-ruler, as both the EW penetration depth and the relative emission into under- and supercritical angles varies over this length scale - thinner dye layers, more similar to the TPPS layer but spectrally distinct and brighter, would be advantageous;

(*iv*), controlling the final dye concentration, the degree of surface coverage and spatial distribution depends - for the used fluorophore pair - on aggregate formation (for TPPS) and on crystallisation (for Rubpy), as well as on the very surface chemistry. Changing, e.g., from a BK-7 substrate to a quartz coverslip is therefore not trivial. A more standardised and modular approach would be preferable and generate flexibility for using other colours, different substrates and providing a more uniform emitter layer thicknesses.

Finally, (*v*) other fluorophore distances and more colours are possible[43], and while we present in this paper an application on the length-scale of *λ*/5, larger fluorohore-spacings are possible, opening up applications in very different fields of application, ranging from standards in FRET or MIET assays, up to axial metrology on much larger (μm) scales in confocal, two-photon or light-sheet microscopies.

## 5. Conclusion

Altogether, our experimental results show that color-multiplexed nano-sandwiches are a promising avenue for benchmarking optical confinement and they allow for a better quantitative interpretation of TIRF-intensity data. Importantly, even without the precise knowledge of the EW decay, the presented dual-color sandwiches can be used in a semi-quantitative manner - as shown here - to quantify the relative "absence" of the more distant colour from their respective colour channel (and using the proximal fluorophore as an intrinsic control). Alternatively - when a more quantitative interpretation is sought for - the user can record the mixed spectrum and unmix the relative fluorophore contribution to generate a continuous figure-of-merit.

**Acknowledgements**

This study was financed by the European Union (H2020 Eureka! Eurostars, "NanoScale," E!12848, https://nanoscale.sppin.fr, (to M.O. and A.S.), with co-funding from the Centre National de la Recherche Scientifique (CNRS), University Paris Cité, and BINA Nanocenter. The authors are grateful for support from a Franco-Israeli Maimonide grant, "NanoCoq" as well as the France-Bio-Imaging large-scale national infrastructure initiative (FBI, ANR-10-INSB-04, Investments for the future). The Oheim lab is a member of the C'Nano Excellence Network in Nanobiophotonics (CNRS GDR2972).

**Supporting Information**

Supporting Information is available online.

**Conflict of Interest**

The authors declare no conflict of interest.

**Author Contributions**

M.O. and A.S. supervised, conceived, and designed this research. M.F. and I.O. fabricated samples, performed measurements, and analyses. Y.A. conceived the $MgF_2$ samples; M.O. and O.S. conceived and designed the microscope and image acquisition and analysis software. O.S. and Y.A. programmed the image-analysis tools. All imaging experiments were done by I. O., M.O., and A.S. All authors contributed to the interpretation of data and the writing of the manuscript.

# SUPPORTING INFORMATION ONLINE

A Colour-Encoded Nanometric Ruler for Axial Super-Resolution Microscopies£

Ilya Olevsko, Omer Shavit, Moshe Feldberg, Yossi Abulafia, Adi Salomon*,✉ and Martin Oheim*

* co-last authors

✉ adi.salomon@biu.ac.il

This supporting information contains

## List of used abbreviations

| | | |
|---|---|---|
| AF | - | autofluorescence |
| AFM | - | atomic force microscopy |
| BFP | - | back-focal plane |
| CT | - | corona plasma treatment |
| DIC | - | differential interference contrast |
| DIW | - | de-ionised water |
| eBeam | - | electron beam |
| EW | - | evanescent wave |
| FTM | - | film-thickness measurement |
| LBL | - | layer-by-layer |
| MW | - | molecular weight |
| NA | - | numerical aperture |
| PDDA | - | Poly-diallyl-dimethylammonium-chloride |
| SAF | - | supercritical angle fluorescence |
| RI | - | refractive index |
| Rubpy | - | Tris (bipyridine) ruthenium (II) chloride ($[Ru(bpy)_3]^{2+}$ $2Cl^-$) |
| SIM | - | structured-illumination microscopy |
| SMLM | - | single-molecule localisation microscopy |
| TEM | - | transmission electron microscopy |
| TIRF | - | total-internal reflection fluorescence |
| TPPS | - | 4,4′,4″,4‴-Porphine-5,10,15,20-tetrayl tetrakis benzene-s sulfonic acid |
| UAF | - | undercritical angle fluorescence |

**SUPPORTING FIGURES**

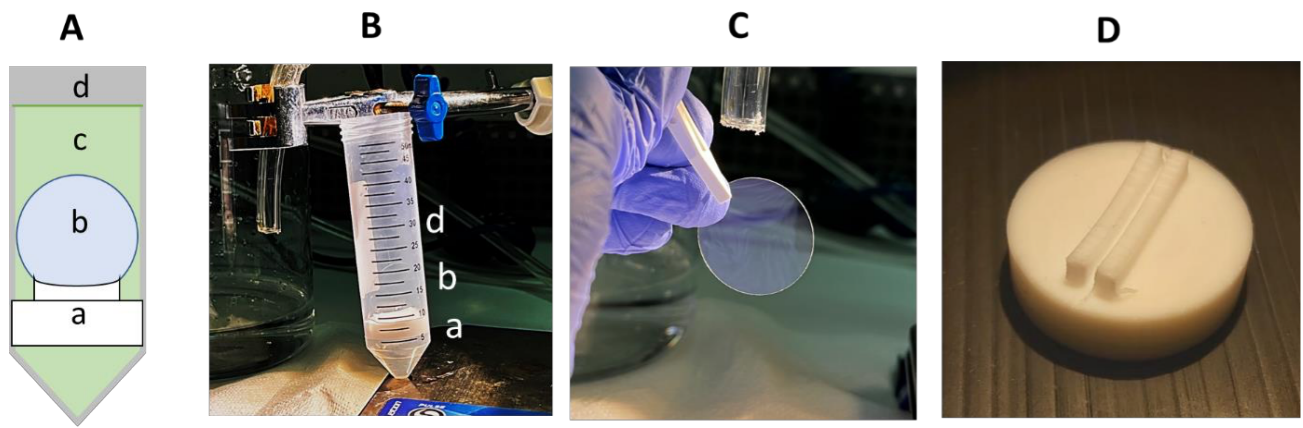

**Figure S1**. *Cleaning procedure for the used borosilicate substrates.* (A), schematic illustration of the substate cleaning chamber: *a,* coverslip-glass holder, made from teflon. *b*, mounted 170-µm thick, 25-mm diameter glass coverslip (substrate). *c*, Hellmanex cleaning solution. *d*, 50-ml Flacon plastic tube. (B), photo of the cleaning chamber, without solution. *a*, *b*, and *d* identify the elements as shown in A. (C), photo of a substrate, after cleaning. We used the reflection to reveal surface impurities, if any. (D), photo of the custom teflon holder. Cleaned coverslips were stored in a clean and dry Falcon tube after this procedure, prior to use.

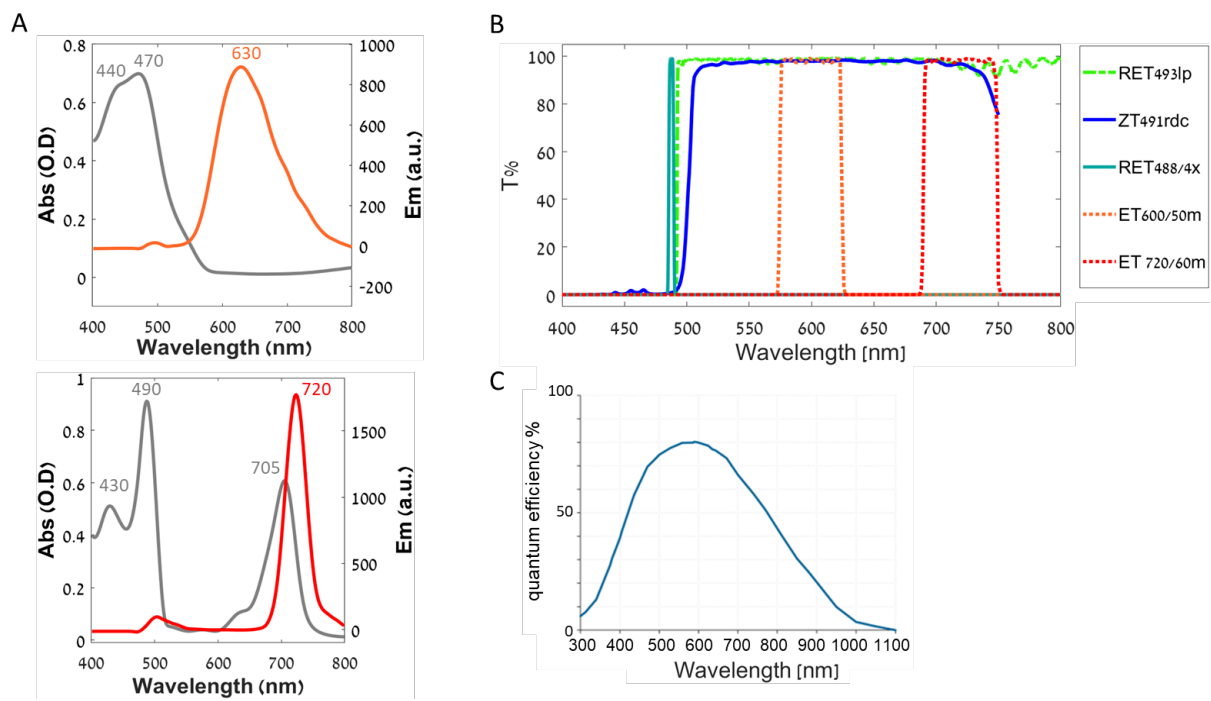

**Figure S2**. *Dye, filter and detector spectra.* (A) Absorption (*gray*) and fluorescence emission (*colored*) spectra of the used dye solutions. *Top*, 1 mM Rubpy aqueous solution (*orange*) . *Bottom*: 1 mM TPPS J-aggregate in acid (pH=1), water as a solvent (*red*). All spectra were measured from a drop of solution sandwiched between 2 cover slips and using an inverted microscope (IX83) linked to a photospectrometer. (B), transmission spectra of the used laser clean-up filter (*turquoise*), dichroic mirror (*blue*), long-pass (*green, dash-dotted*) and emission band-pass filters (orange and red dotted lines, respectively). Curves are replotted from the respective supplier data. (C), quantum efficiency of the used sCMOS camera (adopted from the PCO website).

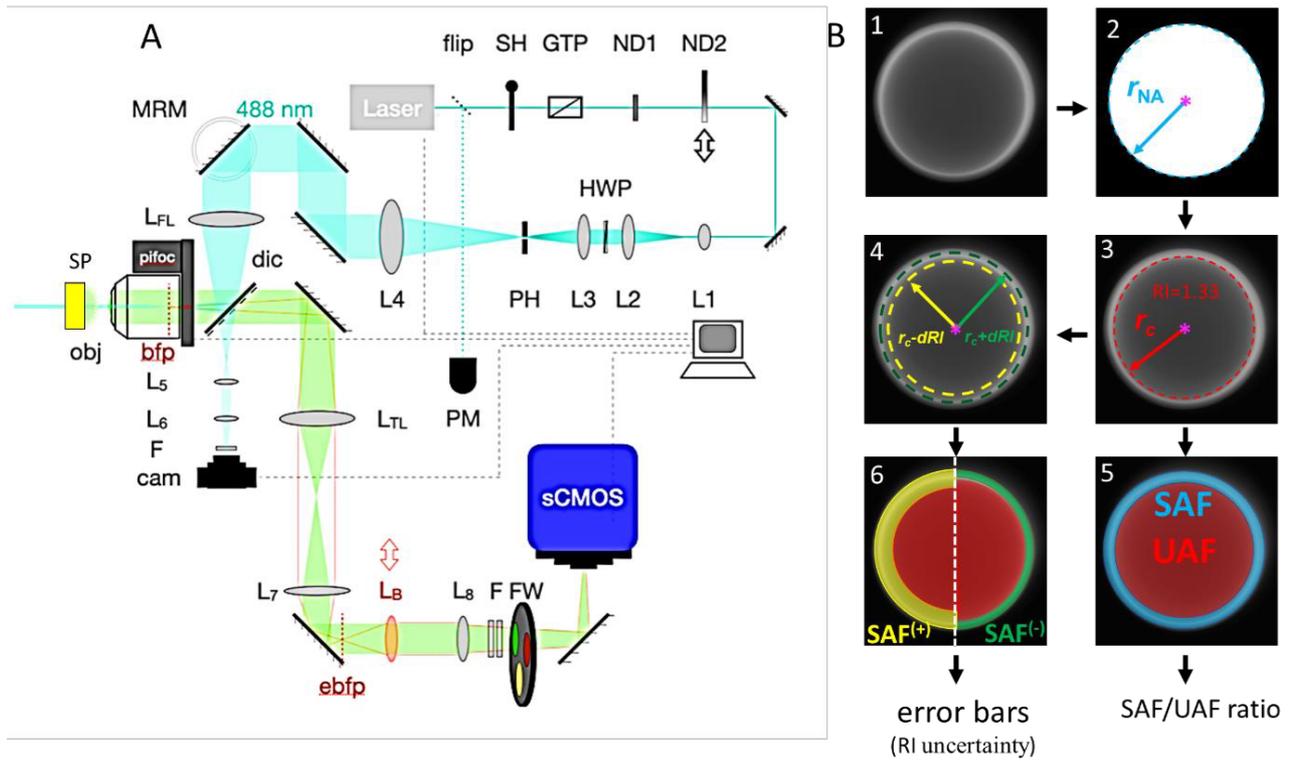

**Figure S3**: *SAF Setup and BFP-image analysis*. (A), optical layout of the used variable angle TIR-SAF microscope. *Turquoise*: excitation optical path. Laser (488 nm), flip - flippable mirror; SH - shutter; GTP - Glan-Thompson polarizer; ND - neutral- density filters; L1, L2, – first Kepplerian beam expander; HWP-half-wave plate; L3, L4 - 2$^{nd}$ beam expander, PH – pinhole (spatial filter); MRM - mechanical rotation mirror - to alter the polar and azimuthal beam angle in the sample plane to get TIRF or EPI imaging; $L_{FL}$ - focusing lens; dic - dichroic mirror; pifoc - piezo-electric focus drive; obj – ×100/1.46NA microscope objective; bfp- backfocal plane; SP- sample plane. *Green*, emission optical path: fluorescence is collected through the same obj, passes the dic and is reflected towards $L_{TL}$ - tube lens; L7, L8 – beam expander; ebfp - equivalent (conjugate) back-focal plane, $L_B$ - motorized Bertrand lens (toggles between SP and BFP); F- stationary emission filters (2× 500-nm long-pass); FW - motorized filter wheel with emission filters, see *Fig. S2B*; sCMOS - scientific complementary metal-oxide sensor (detector). (B), *1-6*, BFP-image analysis steps: *1*, the BFP image is background-subtracted using a dark image. *2*, after thresholding, the radius of the bright disk identifies the limiting radius, i.e. the effective NA ($r_{NA}$). *3*, a first-guess RI of the sample is provided by the user (shown here, water: 1.33). The radius corresponding to the emission critical angle ($r_c$) was calculated as $r_c = f_{obj} \cdot$ RI, where $f_{obj}$ = 1.65 mm. *4*. dRI adds an error bar (d$r_c$) accounting for RI uncertainty. *5*, the BFP is segmented into 3 regions, SAF, UAF and background. Intensity integration over the SAF (*blue*) and UAF areas (*red*) precedes calculating the SAF/UAF intensity ratio *R*. *6*. Integration over the smaller/bigger ±d$r_c$ areas results in 2 alternative regions (*yellow, green*), generating error bars for the intensity ratio *R* (see Klimovsky *et al.* 2023[1] for details).

Color-coding axial distances

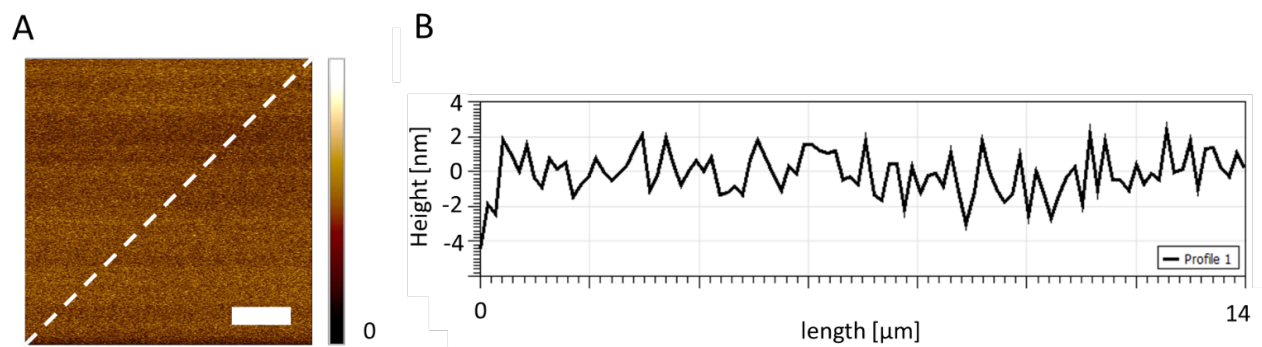

**Figure S4**: *AFM-measured surface roughness of the bare MgF$_2$ substrate.* (A), example atomic force microscopy (AFP) image of a 10 μm by 10 μm bare MgF$_2$ surface area, scanned after corona plasma tretment (CT). The average roughness ($R_a$) was 1.4 nm. Scale bar is 2.5 μm. (B), Height profile (dashed line of 14-μm length) across the MgF$_2$ layer presented in panel A.

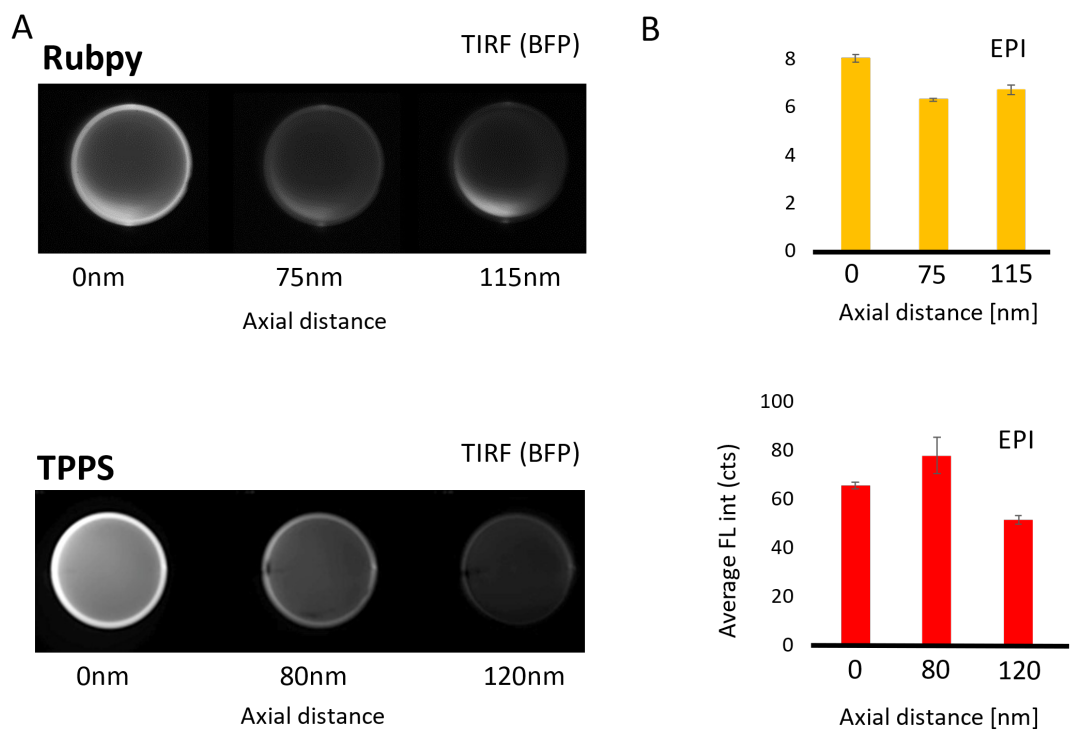

**Figure S5** *Distance-dependance of the emission from TPPS & Rubpy nanometric axial rulers.* (A), *upper images,* representative back-focal plane (BFP) fluorescence images upon evanescent-wave excitation (TIRF) of a series of three Rubpy layers directly on the BK-7 substrate (0 nm) or else at 75 or 115 nm above, separated by a $MgF_2$ Ebeam-deposited spacer layer. *Lower row* shows same, for a TPPS j-aggregate axial series, fluorophore heights as indicated . (B) Epifluorescence intensity measured from these samples (average and SD from 3 areas on each sample) and seen through the respective band-pass filters 600/50M for Rupby and 720/60M for TPPS respectively. In either case, the samples were exited with a 488-nm layer beam. Note the relatively uniform emission upon EPI exception vs. the marked drop in overall intensity and in the intensity of the outer SAF ring upon TIRF.

## SUPPORTING EXPERIMENTAL PROCEDURES

*Substrate cleaning.* Optical-grade borosilicate (BK-7) cover slips (170-μm thick, 25-mm diameter, Menzel-Gläser) were used as substrates for sample preparation. The first step was the removal of any contamination from their surface. To this end, the coverslips were mounted on a custom teflon holder that exposed optimally the surface to the cleaning solution (see Fig. S1). The holders were custom fabricated (CNC). The substrates snuggly fitted into a 50-ml Falcon tube filled with 5% Hellmanex®(III) solution: 1/20 in doubly-distilled de-ionised water (DIW, 18.2 mΩ). The tube was placed in a sonicator (15 min, 50°C). Then, the solution was replaced with 1% Hellmanex®(III) solution, sonicated again (15 min, 50°C) and finally the same was repeated with DIY only. Next, the substrates were removed from the holder, washed in ethanol (99%) followed by DIW and dried under air stream (1 min). The reflection from the surface of the glass substrate was examined under bright light to reveal any contaminations were left. Clean samples were transferred to sterile and dry Flacon tubes until use.

*Stylus Profilometry.* We used a commercial profilometer (DektakXT 150, Bruker) to measure the actual thickness of the deposited $MgF_2$-spacer layers. A line-scan was performed in the 'Hills and valleys' mode along a 2-mm line, 10 s scan time, 3 mg stylus force, 6.5μm stylus *z*-axis range. We scanned both uncovered and $MgF_2$ layers. The height difference of the step determined the layer thickness. 5-6 line scans were preformed and averaged. The standard deviation (SD) of the results was used as the axial position error.

*Ellipsometry* was used to measure the refractive index (RI) of the thin $MgF_2$ layers and provide a reference for roughness and thickness of the deposited layers, to. The technique is based on changes in the polarization of the incident light which is back reflected from the surface. In general, one can use this method for determine the sample composition and phases , thickness, roughness and grading and RI from fitting the optical response from the sample to an assumed material-dependent model. We used a J.A. Woollam ellipsometer in which a polychromatic linear polarized incident light beam (280-1600 nm) impinging at the specimen at 3 different angles (55°, 60°, and 65°, respectively) and back-reflected detection. A mathematical model taking into account the optical responses of the studied material structure and intrinsic parameters was fitted with the measured polarisation state of the reflected light and a low mean-squared-error (MSE) value taken as an indicator for a good fit.. The $MgF_2E$ model (based on experimental measurements of pure crystalline $MgF_2$ from the '*Handbook of Optical Constants*' by Edward D. Palik [2] was used The MgF$_2$ layers were either deposited on a Si wafer with 1-μm SiO$_2$, or else by using a milky white tape onto the back side of the sample to prevent back-side reflections.

*Atomic force microscopie (AFM).* We determined the he average roughness ($R_a$),

$$Ra = \frac{1}{l_r} \int_0^{l_r} |z(x)|\, dx$$

of the deposited spacer and dye layers by AFM (Bio FastScan AFM machine -Bruker AXS). Its axial resolution (<1 nm) allowed us identifying the surface texture and roughness. Measurements were performed in the 'soft-tapping' mode using a silicon tip with silicon-nitride cantilever (Brucker FASTSCAB-B model $T$ = 0.3 μm, $L$ = 30 μm, $k$ = 1,8 N/m, $f$ = 450 kHz). Thickness and roughness analyses provided us with information about the topography differences in the scanned surface area, and it provided $R_a$; $z_{max}$ (the highest point of the surface); the area difference (in %) i.e., the deviation of the heights from the mean value. The lower all three of these values are, the smoother is the surface. We also measured the 'adhesion' parameter. This identifies the presence of different materials on the surface and was used to characterize the homogeneity of the dyes layers ib the MgF$_2$ layers. AFM image analysis was performed using Gwyddion with 2nd degree polynomial leveling.

*Fabrication of single-colour Rubpy axial rulers.* We fabricated a series of calibration slides by depositing them on MgF$_2$ spacers with different heights, similar to what we reported for TPPS dye-layer and MY-133-MC polymer spacer layers in Klimovsky et al. (2023)[1]. A series of mono-dye samples were created according to the fabrication and characterization steps described in the main text. These samples gave the results presented in *Fig. S5*. The SAF/UAF intensity ratios measured from BFP image showed the expected exponential decay [1,3,4]. Yet, the sample series suffered from two major problems: (*i*), differences in the dye layer among different samples of the series compromised the accuracy of the wanted axial calibration. As every sample was coloured separately, small diversities in the deposition process lead to slight concentration and texture differences, which are appreciated on EPI images. Even though the SAF/UAF intensity ratio analysis cancels out variables linearly affecting the intensity (concentration, illumination intensity, detector offset) we noted differences and outliers for some samples of the axial ruler series. (*ii*), a different problem arose from the need to change samples between measurements during the axial calibration measurement. Slight focus and tilt errors as well as the appearance of air bubbles in the immersion oil can result in slightly different conditions between TIRF measurements that adversely affect the accuracy. Also, the change of samples was time consuming and complicated the calibration process. Altogether, these drawbacks prompted us to fabricate the single-excitation, dual-emission sandwiches presented in this study.

*Sectored samples: different fluorophore heights on one coverslip*. The mentioned complications led us to develop as an intermediate step the sectored spacer layout (Fig. S5*A*), in which the different spacers were deposited sequentially on the same substrate, resulting in several flurophore heights side by side on the same test sample. As a consequence, fabrication differences are minimized, and the navigation between different fluorophore heights does neither require a sample change, nor long translation distances. For the sample shown in *Fig.S5A*, the target heights were, respectively, 0 nm (bare substrate surface), 75 and 115 nm. Measured heights (determined by stylus profilometry) of the $MgF_2$ spacers were very close and were 75.2 ± 3 nm, 112.8 ± 4 nm), very close to the target values. Rubpy thin films were then deposited simultaneously on all the spacers using the drop LBL technique as described (10 μl of a 10-mM solution). After drying in a low-pressure chamber, a MY-133-MC cover layer was applied on the entire surface.

Epifluorescence measurements revealed a similar intensity emitted from different sample regions (Fig. S5*B*). In contrast, and as expected, when switching to EW excitation, TIRF images showed continuous intensity drop with increasing fluorophore height. We also calculated from BFP images the more robust SAF/UAF intensity ratio, *R* (see Klimovsky et al. 2023 [1] for details). Although the number of points is too low (3) on this particular sample for a reasonable fit, the observed drop is compatible with the exponential decay (Fig. S5*C*). The *R* values obtained from different areas within each segment (different spacer heights) varied <4%, corroborating the fabrication of a uniform, smooth and homogenous test sample.

*Rubpy & TPPS single-excitation, dual-emitter sandwiches*. We therefore turned to the fabrication of multiplexed dual-dye sandwiches, consisting of stacked Rubpy and TPPS aggregate film. To this end, we first needed to find conditions in which both layers emitted with similar intensities (upon EPI). This was not an easy task, mostly due to intrinsic differences in fluorophore brightness and due to their decrease in intensity once deposited as a film. Also, the spacer deposition on top of them further modulated the brightness of the films (bout 20% for TPPS, about 90% (!) for Rubpy, *data not shown*).

The here presented multiplex samples were fabricated using the following parameters:

(*i*), TPPS-Rubpy multiplex (i.e., Rubpy above TPPS layer): 4 layers of TPPS J-aggs were deposited one after the other on the glass using drop LBL without plasma treatment (PDDA→blow→TPPS→blow→repeat ×4). To minimize re-dissolving, 0.5 mM, pH=1.2 TPPS solution and PDDA were used. After $MgF_2$ spacer deposition, the surface was again plasma-treated, and Rubpy was deposited to form a thin crystalline layer using 1mM solution.

(*ii*), Rubpy-TPPS Multiplex (TPPS film above Rubpy). The Rubpy layer was deposited on the borosilicate substrate by drop LBL following air plasma treatment and using 10-mM solution (i.e., the same conditions as for single-layer Rubpy samples). Then, the $MgF_2$ was applied, and the TPPS layer deposited on the spacer using PDDA solution and plasma treatment for the drop LBL technique (0.5 mM, pH=1.2 TPPS solution)

Fabrication of thin films can be sometimes more art than science, but with the here reported parameters, the resulting layers showed reproducible and bright fluorescence over large areas. Small modifications in the LBL deposition, significantly improved the FL balance between the different emitting films. The designed spacer thickness was 90nm (corresponding to 87 4 ±nm determined by profilometry) for either permutation of TPPS and Rubpy. We deposited the spacer was deposited all over the first dye layer, using very gentle conditions to form a stable $MgF_2$ layer with minimal damage to the dye layer underneath. The main disadvantage of the here used Rubpy films for axial calibration samples is their relatively large thickness (≈30 nm compared to 4-5 nm for TPPS). In addition, neither TPPS nor Rubpy are not commonly used in biology (in fact, neither of the two dyes is biocompatible when in solution), so the difference between their emission spectra in the and the marker(s) used for labelling the biological specimen can cause an interpretation error. In the future it will be interesting to develop a modular way of depositing dye layers that is compatible with standard biological fluorohores.

Notwithstanding these limitations, our work clearly demonstrates our capacity of producing controlled, uniform and bright dual-colour sandwich calibration samples that will be useful for axial metrology in microscopy, but also in related thin-film applications.

## SUPPLEMENTARY REFERENCES